\documentstyle[aps,pre,epsfig]{revtex}

\newcommand{\be}{\begin{eqnarray}}
\newcommand{\ee}{\end{eqnarray}}
\newcommand{\beq}{\begin{equation}}
\newcommand{\eeq}{\end{equation}}
\newcommand{\xx}{\begin{eqnarray*}}
\newcommand{\yy}{\end{eqnarray*}}

\begin{document}
\title{Relevance of soft modes for order parameter fluctuations \\
 in the Two-Dimensional $XY$ model}

\vskip 3.0cm

\author{
B. Portelli and P.C.W. Holdsworth}

\address{Laboratoire de Physique, Ecole Normale Sup\'erieure de Lyon, 46
 All\'ee
d'Italie, F-69364 Lyon cedex 07, France
}                           
\maketitle
\vskip 1cm

\begin{abstract}

We analyse
the spin wave approximation for the 2D-XY model, directly in 
reciprocal space. In this limit the model is diagonal and the normal modes
are statistically independent. Despite this simplicity non-trivial critical
properties are observed and exploited. We confirm that the observed asymmetry
for the probability density function for order parameter fluctuations
comes from the divergence of the mode amplitudes across the Brillouin zone.
We show that the asymmetry is a many body effect despite the importance
played by the zone centre. The precise form of the function is dependent on the
details of the Gibbs measure giving weight to the idea that an effective Gibbs
measure should exist in non-equilibrium systems if a similar distribution is
observed.

\end{abstract}

\vskip 2cm

\section{Introduction}

Order parameter fluctuations in critical systems are known to be
non-Gaussian~\cite{wil.74} and the form
and universal nature of the fluctuations  has
 been studied from a field theoretic and
renormalization group point of view~\cite{Bruce,Binder,chen,botet}. 
More recently a surprising and more pragmatic motivation has come from
the experimental observation
that the probability density for fluctuations of global measures in correlated,
but non-equilibrium systems, such as energy injected into a closed 
turbulent flow, or global measures of self-organized criticality can 
be of the same form as those for order
parameter fluctuations in a critical system~\cite{BHP,prl,pin.99,aji.01}.
In this context we have made a series of studies of order parameter
fluctuations in the low temperature phase of the 2D-XY 
model~\cite{JPA,PRE,field}. This is arguably the simplest system in which
one can study critical phenomena as, in zero field a continuous line of
critical points stretches from the Kosterlitz-Thouless-Berezinskii
phase transition~\cite{KT}, $T_{KT}$, to zero temperature. 
Critical behaviour therefore occurs
even at temperatures where a harmonic approximation to the starting
Hamiltonian is valid. Further, it follows from a renormalization group 
analysis~\cite{JKKN} that the behaviour at all temperatures below $T_{KT}$
is perfectly captured at large length scales by a harmonic Hamiltonian
with effective coupling. As a result, we have been able to calculate
the probability density function $P(m)$ for all temperatures below which
vortices make a measurable contribution, without either invoking the scaling
hypothesis or using renormalization group theory~\cite{PRE,field}.

In this paper we take further advantage of the simplicity of the model,
making a study of $P(m)$ directly in Fourier space. The harmonic Hamiltonian
is diagonalized in reciprocal space and the problem
becomes one of a set of statistically independent normal modes. One can thus
study the contribution to $P(m)$ from different parts of the
Brillouin zone. In section II we show that it is essentially the soft modes near 
the zone centre that are responsible for the asymmetric non-Gaussian 
form for $P(m)$. However, we show further that
a complete and detailed  description of the $P(m)$ requires many modes
and the problem remains intrinsically a many body one.

In section III we discuss our results in relation to extremal statistics. 
We have previously fitted $P(m)$ to a function reminiscent of Gumbel's 
first asymptote from extremal statistics. With this description in terms of
  normal modes
we are in a position to study the contribution from the mode with 
the largest amplitude in each configuration. We confirm that it is not 
sufficient
to keep the largest one of the statistically independent modes and conclude
 that,
if extremal statistics are relevant for the description of the global measure,
then they must be applied to composite correlated structures and not
to the most fundamental excitations of the problem.

In section IV we study the effect of changing the form of the microscopic
generator for the normal modes. We show that changing from the Gibbs measure 
to an arbitrary form does change the final outcome for the global fluctuations.
We generalize the Gibbs measure analytically using a $\chi^2(\nu)$ distribution
for the contribution from each mode. This function gives the probability for
a composite variable constituting a sum of $\nu$ statistically independent 
degrees of freedom. We show that the global measure becomes Gaussian as $\nu$
becomes large and give a physical interpretation for this. Some conclusions
are presented in section V.


\section{Description of the Two-Dimensional $XY$ model in Fourier space}
 
\subsection{Framework of the study}
In the $2D-XY$, or plane rotator model classical spins
of unit length, confined to a plane in spin-configurational space
interact via the Hamiltonian
\begin{equation}
H=-J\sum_{<i,j>}\vec{S}_{i}\cdot\vec{S}_{j}=-J\sum_{<i,j>}\cos(\theta_{i}-
\theta_{j}),
\end{equation}
where the orientation of the spin vector is give by the angle $\theta_i$.
$J$ is the exchange constant, 
the sum is over nearest-neighbours and here we consider spins on sites
of a square lattice with periodic boundaries and lattice constant of
unit length.

At low temperature the harmonic, or spin wave Hamiltonian
can be written 
\begin{equation}\label{H-lin}
H={1\over{2}}\sum_{BZ}J\gamma_{q}{\varphi_{q}}^{2}
\end{equation}
where  $\phi_{q} = Re\left[ 1/\sqrt{ N} \sum_i \theta_i \exp(-i\vec q.
\vec r_i)\right]$ and $\gamma_{q}=4-2\cos(q_{x})-2\cos(q_{y})$.
Here $\phi_q$ takes on values between $\pm \infty$ as we have neglected
the periodicity of the original spin variables.
The sum is over the $N-1$ modes of the Brillouin zone (BZ)
neglecting the Goldstone mode at $q=0$. The latter is a purely 
diffusive mode contributing zero in energy. As discussed below we find that we can 
replace $\gamma_q = q^2$ without introducing an error in the calculated distribution
function.

In principle this Hamiltonian correctly describes the critical
behaviour throughout the low temperature phase as bound vortex pairs
are renormalized away at very large length scale. In practice this
is true for the finite system sizes studied here only  as long
as no vortex pairs are excited. However, as vortices only appear
in a relatively small range of temperature close to the $KTB$ 
transition~\cite{web} there remains a large range of temperature
over which the behaviour is correctly described by a Hamiltonian of
this form. Within this regime non-linear but analytic terms in
the expansion of $\cos(\theta_{i}- \theta_{j})$ are accounted for
perfectly by an effective coupling constant~\cite{JKKN,PRE}.



Thus, at low temperature, the $2D$ $XY$ model can be viewed as a polydisperse 
perfect gas of particles of ``mass'' $J\gamma_q$.
The probability function for each mode is therefore a
Maxwell-Boltzmann distribution
\begin{equation}\label{micro-p}
P(\varphi_{q})=
\sqrt\frac{\beta J\gamma_{q}}{2\pi}
\exp(-{\beta J\gamma_q\varphi_{q}^{2}\over{2}})
\end{equation} 
depending explicitly on the vector $\vec q$.

The aim of this article is to show 
that this dispersion of mass, associated with a Gibbs measure is at the heart
 of the mechanism of criticality
and that it leads directly to the asymmetry of the probability
density function (PDF) for the magnetic order parameter. Given
the separable nature of the problem we are able to do this by
performing numerical simulation and analytic work directly in Fourier
space, without referring to the spins in real space.
The order parameter $m$   for a single configuration    can be defined 
\begin{equation}\label{m}
m=
\frac{1}{N}
\sum_{i=1}^{N}\cos{(\theta_{i}-\bar\theta)}\hspace{0.5cm} 
\end{equation}
where
\begin{equation}
 \tan\bar\theta=\frac{\sum_{i}\sin\theta_{i}}{\sum_{i}\cos\theta_{i}}
\end{equation}
is the instantaneous magnetization direction. With this definition,
$m$ is a scalar quantity giving the length of the
magnetization vector independently of its direction in the spin plane.

In a previous paper~\cite{PRE}, we have calculated the PDF 
for a large, but finite size system, as a function of the reduced variable
$\mu = (m-\langle m\rangle)/\sigma$, where $\langle m\rangle$ and $\sigma$ are 
the mean and the standard deviation of the distribution:
\be\label{PDF}
P_L(\mu)&=&\int_{-\infty}^{\infty}\frac{dx}{2\pi}e^{ix\mu}\psi(x) \\ \nonumber
\ln{\psi(x)} & =&
-i x\sqrt{\frac{1}{2g_{2}}}
\sum_{BZ} \frac{G({\bf q})}{N}  
-\frac{1}{2}\sum_{BZ} 
 \ln\left[1-i\sqrt{\frac{2}{g_2}}{G({\bf q})\over{N}}x\right]
\ee
where $P_L(\mu) = \sigma P(m)$, $G({\bf q})=\gamma^{-1}_q$ and 
$g_k= \sum_{BZ} G({\bf q})^k/N^k$. 
The sum over ${\bf q}$ and the integral over $x$, in~(\ref{PDF}) 
are performed numerically and $P_L(\mu)$ approaches a (thermodynamic ) limit
function rapidly for $N > 100$ spins. Surprisingly
this function, shown in Figure 1 and discussed in detail below,
 is independent of temperature and therefore of critical exponent
$\eta = T/2\pi J$ along the line of critical points.

Expanding (\ref{m}) in small angles, keeping only the leading terms
and Fourier transforming, one can write
\begin{equation}\label{m-lin}
m \approx1-\frac{1}{2N}\sum_{BZ}{\varphi_{q}}^{2}=1-\sum_{BZ}m_{q}
\end{equation}
where $m_{q}=\frac{1}{2N}{\varphi_{q}}^{2}$ has the normalized distribution 
\be\label{micro}
P(m_{ q}) = \sqrt{\beta J \gamma_q N\over{ \pi}}
m_{ q}^{-1/2} \, {\rm e}^{-\beta J N \gamma_q m_{ q}}\;\;.
\ee
While this expansion is strictly valid to first order in $T$ 
and for finite systems only,
we have previously shown that the PDF for the fluctuations in $m$,
given by (\ref{m-lin}) is identical to that for $m$ given by (\ref{m}), 
even outside the range of $T$ for which the expansion is valid. For this
linear development of the order parameter, the critical exponent, $\eta=0$.
There are a number of subtil points associated with this result; most 
notably, the violation of the hyperscaling relation between 
critical exponents if definition (\ref{m-lin}) is used. These
points will not be discussed further here; rather, we refer the reader
to references~\cite{PRE,field,aji.01} where they are discussed in detail.
However the important point for this paper is that
the magnetic fluctuations of the
entire low temperature phase can be characterized by the harmonic Hamiltonian
(\ref{H-lin}) and the linearized order parameter (\ref{m-lin}) and
the problem can be analysed entirely in Fourier space, with statistically
independent harmonic degrees of freedom.

\subsection{Simulations in Fourier's space}

We first test these ideas
by comparing the PDF, as generated by (\ref{PDF}), 
against Monte Carlo simulation performed directly in ${\bf q}$-space.
 We take the real part of 
the Fourier transformed spin variable as the independent degree of freedom on
each point in the Brillouin zone and define the quarter of the zone with $q_x$
and $q_y$ positive. Points with $q_x \ne 0$ and $q_y \ne 0$ have a 4-fold
degeneracy, points on the axis, with $q_x=0$ or $q_y=0$ 
have a 2-fold degeneracy
and the origin is excluded, giving the total of $N-1$ points.
For each configuration,
amplitude $\varphi_q$ is generated using (\ref{micro-p}).
The results of the simulation are compared in Figure 1 with the theoretical
result given by equation (\ref{PDF}). Data are shown for $10^8$ configurations.
Agreement is quite excellent, showing clearly that the model of statistically
independent variables and the calculation leading to (\ref{PDF}) are consistent
and that they correctly describe the fluctuations in the low temperature
domain. The PDF is
asymmetric, with a quasi-exponential tail for fluctuations below the mean, 
and with a much faster double exponential fall off for fluctuations 
above the mean~\cite{PRE}. 

\begin{figure}
\begin{center}
\epsfig{file=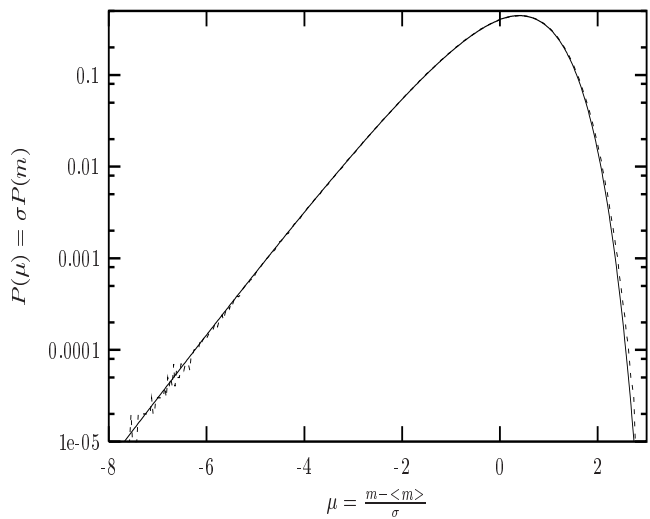,height=7cm,width=9cm}
\end{center}
\caption{The PDF in two dimension ($L=32$)  at temperature $T=1$.
The continuous line is the fast Fourier transform, equation (\ref{PDF}).
The dotted line is for q-space Monte Carlo.} 
\label{simul1}
\end{figure}


\subsection{Relevance of soft modes}

Our aim is now to study the effect on the PDF coming from
different parts of the Brillouin Zone and in particular 
the contribution from the soft modes, near the zone centre.  
We should be able to do this by 
restricting the summations in equation (\ref{PDF}) to the 
desired part of the zone only. We define shells moving out
from the zone centre: the first contains only the four softest modes.
The subsequent shells are squares of length $\frac{2\pi}{L}n$ and unit thickness, 
containing $(2n+1)^{2}-1$ modes. We have tested the PDF generated
from equation (\ref{PDF}), with restricted sums over the shells defined
above, against Monte Carlo data taken for modes in the same shells.
We again found essentially perfect
agreement giving confidence in the theoretical result for both complete
and partial sums over the Brillouin zone. 
In the following analysis we need to make no further reference to
numerical simulation and can restrict ourselves to 
the analytic result of equation (\ref{PDF}). Some initial steps in this 
approach were made in~\cite{rac.94} for the PDF of height fluctuations in
models of interface growth.

Results are shown in Figure 2 for summations over varying numbers of
shells in the Brillouin zone and in Figure 3 for
a calculation including all modes except the four softest modes. One can see
that the asymmetry comes explicitly from the soft modes near the zone
centre. The exponential tail is reproduced to a first approximation with
only the first four modes, while excluding them from the distribution gives
a nearly symmetric distribution.

The reason for this is the dispersion in the typical amplitudes for modes
over the Brillouin zone.  For small $q$ values $\gamma_q \sim q^2$,
which means that the average value, $\langle m_q\rangle$, is of order $O(T/J)$ at the zone centre,
while $\langle m_q\rangle \sim T/NJ$ on the zone edge. The dispersion in contributions
therefore diverges with systems size.
The magnetization is the sum over variables $m_q \sim \varphi_q^2$,
not the $\varphi_q$ themselves and the distribution for the $m_q$, (equation \ref{micro}), 
therefor have exponential
tails.  If terms in the sum are not individually negligible, as is the case here, 
the distribution for the global measure is consequently asymmetric.
However it is not the case that the the global PDF is reproduced quantitatively
by softest modes by themselves. More modes are required for a quantitative fit and one
must include more than the 4  shells shown in Figure 2
for a good  fit by eye. 

The contribution to the PDF from any set of $N_g$ modes  with fixed $q$
is therefore that for an ideal gas of $N_g$ identical particles;
the so called $\chi^2$ distribution
\be
f(x)\sim x^{N_{g}/2-1}\exp{(-\beta x)},
\ee
Writing the $\chi^2$ distribution in terms of $\mu$
one finds
\begin{equation}
P_L(\mu)={\cal N}(s-\mu)^{N_{g}/2-1}\exp{(b(\mu-s))}
\end{equation}
where
\be
 {\cal N}=\Bigl(\frac{\alpha\sigma}{T}\Bigr)^{N_{g}/2}\frac{1}{\Gamma
(\frac{N_{g}}{2})}, \hspace{0.5cm} 
s=\frac{1-<M>}{\sigma}, \hspace{0.5cm} 
b=\frac{\alpha\sigma}{T},\hspace{0.5cm} \alpha=8\pi^{2} J \hspace{0.5cm}
 \sigma=\frac{1}{\alpha}\sqrt{\frac{N_{g}T}{2}}.
\ee
The PDF is therefore a convolution over $\chi^2$ distributions with
varying particle number and ``mass''. Once the discrete nature of the lattice is lost the number
of particles for each value of $q$ is proportional to the density of states,
$n(q) \sim q^{d-1}$. For numerical simplicity we have defined square
shells but our findings are compatible with this continuum description.

We study the goodness of fit for
$n$ shells, in Figure 4,
where we plot the error function
\be
\delta(\mu,N/N_{eff})=\frac{P_L(\mu,N_{eff})}{P_L(\mu,N)}
\ee
with $N_{eff} =(2 n+1)^{2}-1$, for $N=10^4$ sites. The further out in the wings
one goes, the more shells one has to include in order to get a required value of $\delta$.
For large $\mu$ and $N >> N_{eff}$ we find 
\begin{equation}
\delta(\mu,{N_{eff}})\sim 1+{\cal C}(\mu)\frac{1}
{N_{eff}}
\end{equation}
independently on $N$. The many body nature of the distribution is therefore 
confirmed by the fact that $\delta$ only goes to zero as $N_{eff}$ diverges.

We have previously argued~\cite{PRE} that the critical nature of the 2D-XY model is epitomized 
by the fact that all length scales from the microscopic to the macroscopic and
therefore all $q$ values are important. For example, 
calculation of the mean magnetization $\langle m \rangle$ involves an integral of the form
$\int^{\pi}_{2\pi/L} 1/q dq \sim \log(L)$ where both limits of integration come into play.
One could argue that the above result, with $\delta$ becoming independent of $N$
in the limit $N >> N_{eff} >> 1$ is not quite in agreement with this discussion as only
the zone centre is important in this limit. However, exactly the same would be true for
a situation where the central limit theorem is satisfied, that is,
with zero dispersion in the spin wave stiffness of the modes:
here one would build a Gaussian distribution
by adding the modes from different shells and in the above limit, 
this would be attained with arbitrary accuracy by increasing $N_{eff}$ independently of the 
value of $N$. The latter is realised in the presence of a strong magnetic field~\cite{field}
where the dispersion in spin wave stiffness is removed by the finite magnetic correlation
length. The above analysis is not therefore the most sensitive measure of 
criticality, although it does clearly show that the global measure is the result of
many body contributions, while at the same time showing why no error is incurred when replacing
$\gamma_q$ by $q^2$ in the various summations.

\begin{figure}
\begin{center}
\epsfig{file=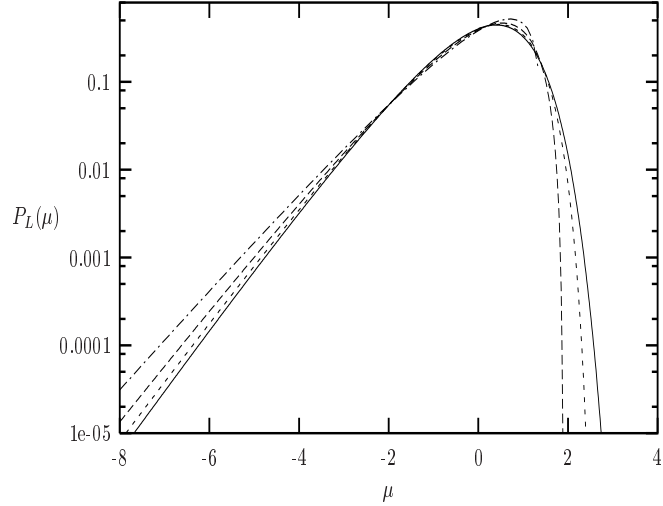,height=7cm,width=9cm}
\end{center}
\caption{PDF generated from the first shell (4 modes, dot-dashed), 1 shells ( 8 modes, long dashed)
2 shells (24 modes, short dashed), full Brillouin zone (solid line). All results are generated from
equation (\ref{PDF}) for $L = 100$}
\label{simul2}
\end{figure}

\begin{figure}
\begin{center}
\epsfig{file=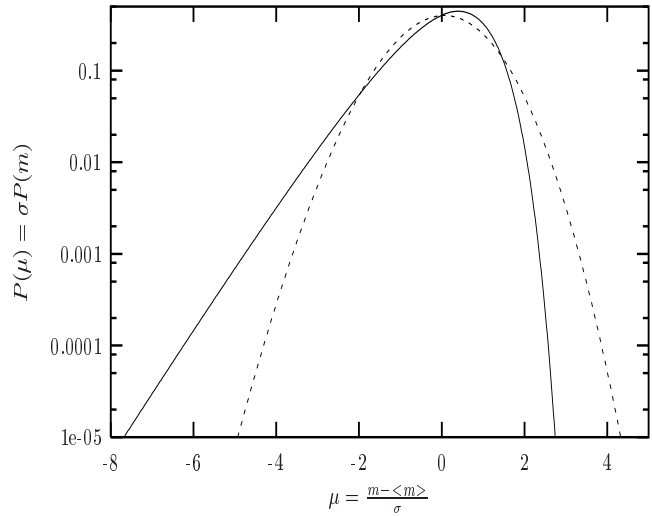,height=7cm,width=9cm}
\end{center}
\caption{PDF generated from all but the first shell ($N-5$ modes, short dashed),  
full Brillouin zone (solid line). All results are generated from
equation (\ref{PDF}) for $L = 100$} 
\label{simul3}
\end{figure}


\begin{figure}[h]
\begin{center}
\includegraphics[width=9cm,height=7cm]{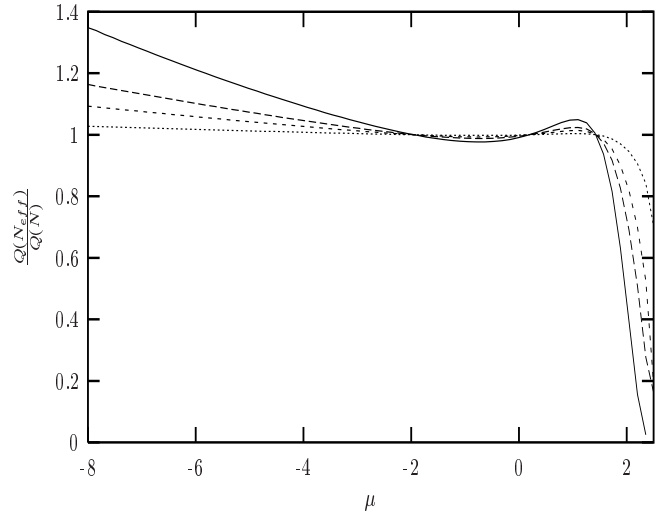}
\end{center}
\caption{$\delta(\theta)=\frac{Q(N_{eff},\theta)}{Q(N,\theta)}$  for a given
 size $N$ and for different $N_{eff}$ ( different fractions ). From top to bottom on the 
left of the figure, the curves are for 4, 8, 25 and 48 modes.} 
\end{figure}



\subsection{Comparison with the three dimensional $XY$ model}

Criticality can be seen more clearly by comparing with the magnetic fluctuations
in the low temperature phase of the three dimensional XY model. Here there is
broken symmetry and long range magnetic order, although the longitudinal
susceptibility remains weakly divergent in the ordered phase and temperature
is a dangerously irrelevant field near $T=0$~\cite{Ma,Maz}. As a result the PDF
for the magnetic fluctuations
is weakly asymmetric~\cite{PRE} as shown in Figure 5. Equation (\ref{PDF}) is
easily  extended to three dimensions and the same partial summations over the Brillouin 
zone can be performed. Included in Figure 5 are the PDF calculated from the softest
modes only and calculated from all but the softest modes. As one can see, the form 
of the curve is well reproduced by the first shell. This should be compared with 
the two-dimensional case where many more modes were required for an equivalent
description. 

\begin{figure}
\begin{center}
\epsfig{file=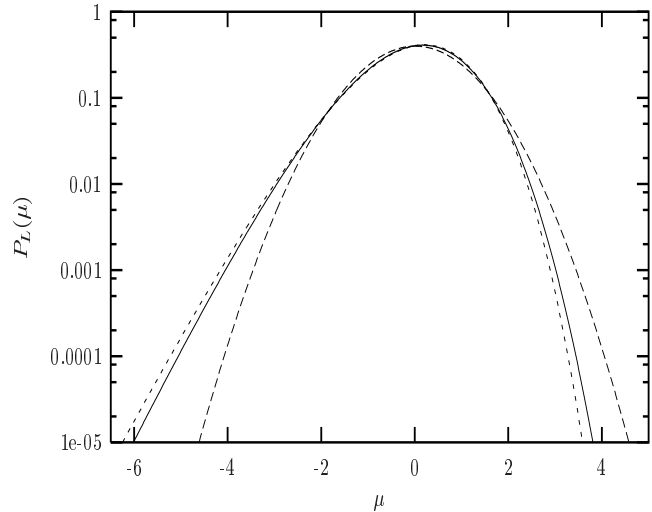,height=7cm,width=9cm}
\end{center}
\caption{PDF for the 3D-XY model for the first shell (6 modes, short dashed), all modes
except the first shell (long dashed), full Brillouin zone (solid line). 
All results are generated from
equation (\ref{PDF}) for $L = 100$} 
\label{simul4}
\end{figure}

This can be seen in more detail by considering the evolution of the
asymmetry of the distribution through the skewness, 
$\gamma$~\cite{PRE},
\be
\gamma = \langle \mu ^3 \rangle = \sqrt{8g_3^2\over{g_2^3}},
\ee
as one consecutively removes shells around the zone centre from the summations
in $g_k$. That is, we now sum over the Brillouin zone except for an ever increasing
number of shells around the zone centre. The results are shown in Figure 6
for two and three dimensions. Clearly it is necessary to remove fewer shells in
three dimensions than in two dimensions, before $\gamma \rightarrow 0$
and the distribution becomes Gaussian.


\begin{figure}[h]
\begin{center}
\epsfig{file=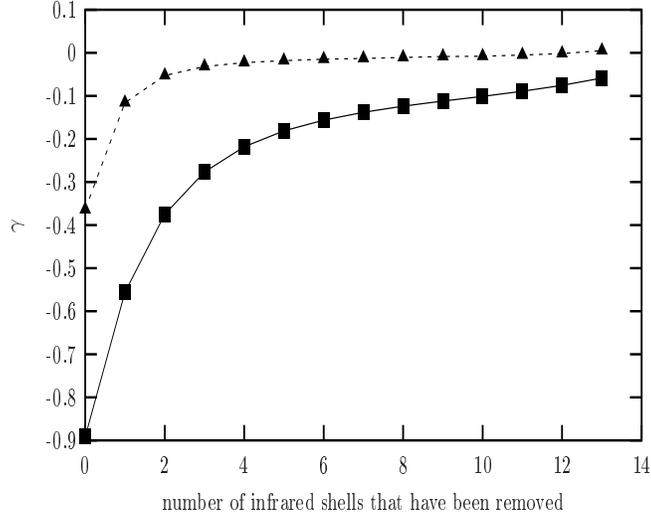,height=7cm,width=9cm}
\end{center}
\caption{Evolution of the skewness $\gamma$ as a function of removed infrared
 shells. Study in two (full line) and three (dashed line) dimensions}
\label{skew} 
\end{figure}

\section{Link with extreme statistics} \label{extreme}

In \cite{prl,PRE}, it was shown that the analytic expression for the
PDF $P_L(\mu)$ is extremely well represented by a function
reminiscent of Gumbel's first asymptote for the statistics of extremes~\cite{gum}
\begin{equation}
P_G(\mu)=w\Bigl(e^{b(\mu-s)-e^{b(\mu-s)}}\Bigr)^{a},
\label{gumbel}
\end{equation}
where 
\be
w & = & {a^a b\over{\Gamma (a)}} \nonumber \\ 
b &= & 
\sqrt{ {1\over{\Gamma (a)}} {\partial^2 \Gamma (a)\over{\partial a^2}}- 
\left[ {1\over{\Gamma (a)}} {\partial \Gamma (a)\over{\partial a}}
\right]^2 } \nonumber \\
s & = &  {1\over{b}} 
\left[\log (a) - {1\over{\Gamma (a)}} {\partial \Gamma(a)\over{\partial a}} 
\right] \,.
\ee
The function therefore has a single independent parameter $a$.
When $a$ is an integer, equation
(\ref{gumbel}) gives the PDF for the $a^{th}$ smallest value from a set
of $N$ random numbers in the limit $N \rightarrow \infty$. 
We have made a comparison between the function $\psi$ of eqn. (\ref{PDF})
and the characteristic function for $P_G$
\be\label{dev-1}
\ln \psi_G(x)&=&\ln\frac{w\Gamma(a)}{sa^a}-ix\left(s+\frac{\Psi(a)}{b}-\frac{\ln(a)}{b}\right)
-\frac{x^2}{2b^2}\Psi'(a)
+i\frac{x^3}{6b^3}\Psi''(a)\\ \nonumber
&+&\frac{x^4}{24b^4}\Psi'''(a)-i\frac{x^5}{120b^5}\Psi^{(4)}(a)+\cdots
\ee
where $\Psi(z)$ is the digamma function $\Gamma'(z)/\Gamma(z)$.
Expanding $\psi$ and equating the first four terms we find an implicit
function
\be\label{implicit}
{\Psi''(a)\over{\Psi'(a)^{3/2}}} = -2^{3/2} {g_3\over{g_2}^{3/2}}.
\ee 
Solving for $a$ and hence for the other constants we find
$a=1.58$, $b=0.93$, $s=0.37$, $w=2.16$. This does not represent the
exact solution for $P_L$: higher order terms in the expansion
of the two functions are not the same and their ratio diverges slowly
from unity with increasing order.
 
Although (\ref{gumbel})
is not the exact solution, mearly a good approximation and the best fit
value $a \approx \pi/2$ is non-integer, it is still interesting to search 
for a concrete link with extremal statistics. As we have reduced our model
to one of statistically independent Gaussian variables the obvious question
seems to be: is the PDF of the dominant mode the same as that of the global measure?
We define a magnetization $m = 1-m_{max}$, where $m_{max}$ is the largest of the $m_q$. 
We compare, in Figure 7 the PDF for this quantity with that for 
the full magnetization. The difference between the two distributions is not
very great, but they are certainly not the same and we confirm what we have seen
in the previous section, that the correct distribution comes from an ensemble of statistically 
independent objects and not a single mode. The extreme values are well described
by the Gumbel function with $a = 1$, despite the dispersion in the
spin wave stiffness. The diverging dispersion means that the largest
mode is always one of only a few modes near the zone centre and it is
by no means clear that Gumbel's asymptote should apply here.

We can conclude therefore that an extremal description is not applicable
to the fundamental excitations in the problem. However, one can
perhaps apply extremal statistics to the complex and correlated 
structures that appear in real space for critical systems and for
correlated systems in general. Indeed, in simulations of the 2D Ising
model at a temperature $T^{\ast}$ slightly below the critical temperature, 
PDFs were found to be very similar to that
for the 2D-XY model, when calculated
for both the magnetization and the magnatization coming uniquely from the 
largest cluster of connected spins~\cite{prl}. 
There are clearly many open questions here
but the above analysis strongly suggests that an observation
of extremal statistics is a consequence of the presence of composite
and correlated objects, rather than being a directly observable 
phenomenon among the most fundamental microscopic excitations that
make up these objects.


\begin{figure}[h]
\begin{center}
\epsfig{file=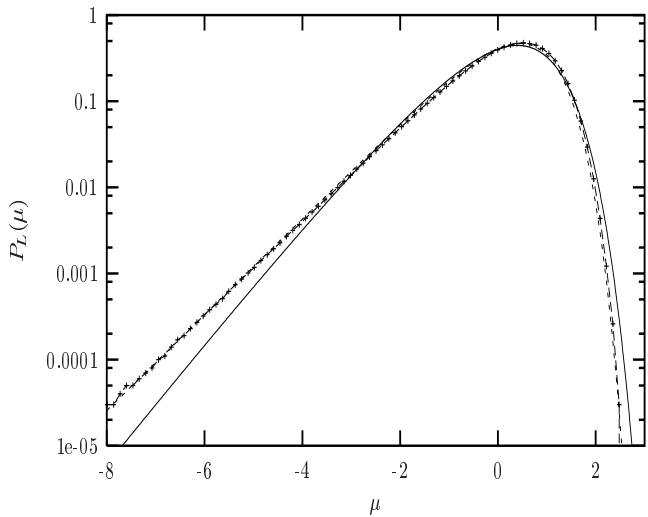,height=7cm,width=9cm}
\end{center}
\caption{Contribution from the Fourier mode of extreme amplitude (dashed line) and comparison
 with the PDF generated from equation (\ref{PDF}) (solid line) and the Gumbel asymptote, $a=1$ (dotted line).}
\label{extr} 
\end{figure}

\section{Influence of the microscopic measure on the global measure}

One might also ask the question: is the global PDF dependent on the existence 
of a Gibbs measure, or effective Gibbs measure for the fundamental excitations
that contribute to the global fluctuations? Within the present framework 
this question can be addressed directly by changing the form of the
generating function (\ref{micro}), while keeping the same dispersion for the standard
deviations, $\sigma_q = \sqrt{\langle m_q^2 \rangle - \langle m_q \rangle^2} = {1\over{\sqrt2 \beta JN\gamma_q}}$. In Figure 8 we show
the global distribution produced from an array of flat generators
\be\label{flat}
p(m_q) & =& 1/  x_q , \;\; m_q \le  x_q ,\\ \nonumber 
p(m_q) & =& 0, \;\; m_q >  x_q ,
\ee
with $x_q = \sqrt{12} \sigma_q$, which satisfies the above criterion.
As can bee seen, the resulting distribution is radically different from
that for the Gibbs measure. It is prefectly symmetric despite the
dispersion in mode amplitudes. The curve is almost, but not quite Gaussian.
We have not made a detailed analysis of this point, but the deviation from Gaussian
behaviour suggests that the central limit theorem is still violated even
though the distribution is now symmetric.
From this exercise we conclude that the dispersion in 
normal modes is not the only criterion for the asymmetric distribution we have observed.
Our results suggest that further to this, we require microscopic generators that are 
themselves asymmetric.
In fact, one can deduce this general result by
considering Gaussian generators for the $m_q$: here, by definition $P_L(m)$
must be Gaussain, as the sum of any set of 
Gaussian variables must itself be a Gaussian variable and therefore symmetric.


\begin{figure}[h]
\begin{center}
\epsfig{file=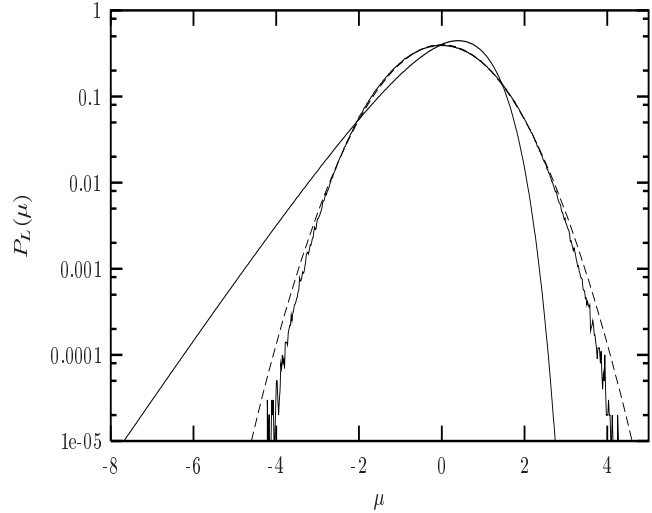,height=7cm,width=9cm}
\end{center}
\caption{PDF calculated numerically from the flat generators, equation (\ref{flat}-symmetric full line)
compared with the PDF from equation (\ref{PDF}-asymmetric full line) and a Gaussian distribution (dashed line). 
All data is for $L=100$.}
\label{flat-fig} 
\end{figure}

One can see analytically how the global PDF varies from the original distribution
to a Gaussian distribution by considering generators of the form
\begin{equation}\label{mqa}
p_{\alpha}(m_{q})=N_{\alpha}m_{q}^{\alpha-1}e^{-m_{q}/ \langle m_q \rangle}.
\end{equation}
Here $\alpha > 0$ in order to ensure the convergence of $p(m_q)$ and
$N_{\alpha}=\frac{1}
{\Gamma(\alpha)m_{0}^{\alpha}}$ is a normalization constant.
We again take $\langle m_q \rangle = 1/\beta JN\gamma_q$ so that 
$p_{\alpha=1/2}(m_{q})$ accounts for the two dimensional $XY$ model with Gibbs measure.

We proceed as previously, defining $m$ within a linear approximation only, for simplicity.
We define
\begin{equation}
P(m)=\int p(m_{q_{1}},...,m_{q_{N}})\delta\Bigl(m-(1-\sum_{q\ne 0}m_{q})\Bigr)
dm_{q_{1}}...dm_{q_{n}}
\end{equation}
and generate  the $m_q$ using equation (\ref{mqa}).
The $\delta$ function imposing the constraints $m=1-\sum_{q} m_{q}$
can be rewritten
\begin{equation}
\delta(m-(1-\sum_{q\ne 0}m_{q}))=\int_{-\infty}^{\infty}\frac{dk}
{2\pi}e^{ik\Bigl(m-(1-\sum_{q\ne 0}m_{q}\Bigr)}.
\end{equation}
As modes are independent the conditional probability
$p(m_{q_{1}},...,m_{q_{N}})$ is just the product
\begin{equation}
p(m_{q_{1}},...,m_{q_{N}})=\prod_{q=1}^{N} p_{\alpha}(m_{q})
\end{equation}
so that 
\begin{equation}
P(m)=\int_{-\infty}^{\infty}\frac{dk}{ 2\pi}e^{ikm}e^{-ik}\prod_{q=1}^{N}
\int_{0}^{\infty}\,dm_{q}p_{\alpha}(m_{q})e^{ikm_{q}}.
\end{equation}
After some straightforward algebra, one can now express the PDF $P_L(\mu)$
\begin{equation}\label{PDFa}
P_L(\mu)=\int_{-\infty}^{\infty}\frac{dx}{2\pi}e^{ix\mu}\psi_{\alpha}(x)
\label{ffta}
\end{equation}
where 
\begin{equation}
\log (\psi_{\alpha}(x))={\Bigl(-ix\sqrt\frac{\alpha}{g_{2}}Tr\frac{G}{N}-\alpha
Tr\log(1-i\frac{x}{\sqrt{\alpha g_{2}}} {G\over{N}})\Bigr)}
\label{psia}
\end{equation}
which is just a generalization of (\ref{PDF}) for arbitrary values of $\alpha$, for which
%
\be
\langle m\rangle =1-\alpha T\sum_{q}\frac{G(q)}{N}, \;\;
\sigma=T\sqrt{\alpha g_{2}}.
\ee
As previously, the expression (\ref{ffta}) can be transformed numerically for large
but finite $N$. This procedure can again be tested by comparison
with Monte Carlo simulation performed directly on the generators $m_{\alpha}$,
The results, which we do not show here, agree as well as they did 
in the initial problem.
In Figure 9 we show $P_L(\mu)$ for various $\alpha$ values calculated from
equation (\ref{psia}). One can see that
increasing $\alpha$ reduces the asymmetry of the global distribution, with it
becoming Gaussian as $\alpha$ becomes large. This result can be interpreted physically
by noting that the generator (\ref{mqa}) is a $\chi^2$ distribution for the
variable $m_q$, which is the sum of $\nu = 2 \alpha$ statistically independent
positive definite variables. One can therefore think of the global PDF as
being constructed from $\nu=2\alpha$ replicas, or identical systems. As $\alpha$
increases, the number of contributions coming from each shell in the 
Brillouin zone, even those near $q=0$ increases. The contribution from each shell
will become Gaussian and $P_L(m)$ will therefore become Gaussian. 


\begin{figure}[h]
\begin{center}
\epsfig{file=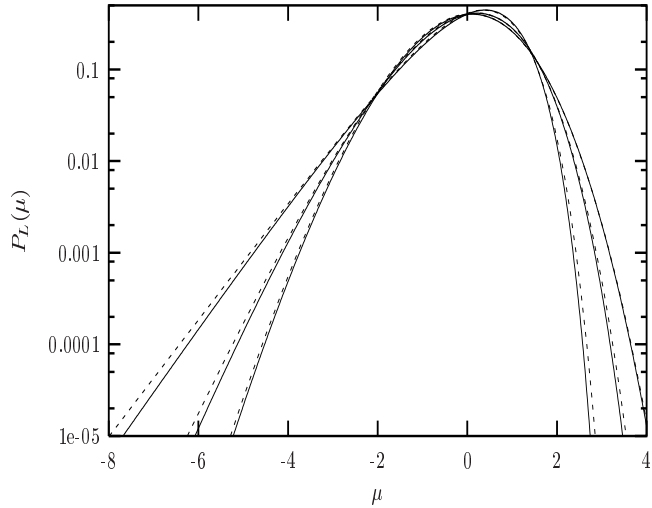,height=7cm,width=9cm}
\end{center}
\caption{Influence of the microscopic distribution $p_{\alpha}(m_{q})$ 
on $P_L(\mu)$, calculated result from equation (\ref{PDFa}-full line) 
generalized Gumbel function fitting function (equation \ref{gumbel}-dashed line).
Data sets for decreasing skewness are for $\alpha = 1/2, 2, 5$.}
\label{alpha} 
\end{figure}

The analysis in terms of the Gumble distribution can also be extended to include
this variation in the form of the microscopic generators. Repeating the analysis
of section \ref{extreme} for the generating function $\psi_{\alpha}$ we find
\begin{equation}
\frac{\Psi''(a)}{\Psi'(a)^{3/2}}=-2\frac{g_{3}}{{g_{2}}^{3/2}}\frac{1}
{\sqrt\alpha}
\label{impl}
\end{equation}
The knowledge of $a$ again allows one to evaluate the other constants $b$, $s$
and $w$
\begin{center}
$b=\sqrt\psi'(a)$ \hspace{0.5cm} $sb=\ln(a)-\psi(a)$ \hspace{0.5cm}
 $w=\frac{a^{a}b}{\Gamma(a)}$
\end{center}
and numerically we find 
\begin{itemize} 
\item[$\bullet$] \underline{$\alpha=2$} :\\
$a=5.45802$, $b=0.44835$, $s=0.21054$, $w=96.563$ 
\item[$\bullet$] \underline{$\alpha=5$} :\\
$a=13.0845$, $b=0.2818$, $s=0.1373$, $w=194537.9$ 
\end{itemize}
Data generated from equation (\ref{gumbel}) is included in Figure 9 
for comparison with the exact solution. Agreement is very good,
indicating once again that the modified Gumbel function is a very useful working
tool for data analysis.
The small deviation between the Gumbel function and equations (\ref{PDFa}),
 observed along the exponential tail
of the distribution for $\alpha = 1/2$ disappears as $\alpha$
increases.

For large values of $\alpha$,  
one can expand (\ref{impl}) using Stirling expansions
for the $\Gamma$ function and its derivatives and find that
\begin{equation}
a\simeq 1+\frac{\alpha g_{2}^{3}}{4g_{3}^{2}}\sim 1+\frac{2\alpha}{\gamma^{2}}.
\end{equation}
We actually find that this expression is quite accurate even outside
the range of $a$ values for which Stirling's formula is strictly valid.

This analysis is very similar to that recently made for
the 2D-XY model in the presence of a magnetic field~\cite{field}.
In that case the field introduces a correlation length, $\xi$, which 
allows one to think of  dividing the system into order $(L/\xi)^2$
statistically independent parts, just as we have proposed above for $\alpha$
different from 1/2.

\section{Conclusion}

  In this paper we have exploited the appealing property of the 2D-XY model, that
non-trivial behaviour of a correlated systems appears despite the model being itself
diagonalizable at low temperature. The magnetization can hence be described in terms of a sum over
statistically independent variables. The PDF for order parameter fluctuations can, in consequence
be written as a convolution over $\chi^2$ distributions for statistically
independent variables, or particles of continuously varying amplitude. We have shown that,
as one might expect for a critical system, it is the softest modes that influence
the distribution most, with their diverging amplitude being directly responsible for 
the violation of the central limit theorem and the observed asymmetry for the PDF
of this global quantity. However, we also explicitly show that the observed asymmetry is
not a single or even a few-body phenomena. Rather, a quantitative reconstruction of the PDF requires the
sum over a number of modes that diverges in the thermodynamic limit. We have also analysed
the ``extreme value'' contribution from the largest mode from each configuration. The
resulting distribution is consistent with the standard theory of extremal statistics
and quantitatively different from the the global PDF. Reconstructing the PDF by summing
successive shells leading out from the zone centre requires an arbitrary number of shells
for arbitrary accuracy. 

We show that the detailed form for the distribution depends on the precise form
for the microscopic distributions for the individual modes. Hence we show that
the form of the global PDF depends on a second criterion, the existence of a Gibbs measure,
and that the nature of the violation of the central limit theorem depends on the
form of microscopic measure. 
This discovery is
perhaps surprising given our empirical observation that a very similar distribution
can be observed for complex systems driven far from equilibrium and in particular that for
injected power fluctuations in a closed turbulent flow at constant Reynolds number~\cite{BHP}.
This extra exigence seems to give weight to the idea that
an effective temperature should exist in such non-equilibrium, steady state systems.
More work in this area is clearly required.

\acknowledgements{ 
It is a pleasure to thank our collaborators S.T. Bramwell, J.-Y. Fortin, S. Peysson, J.-F. Pinton and M. Sellitto
for their contributions to this work and
L. Berthier, L. Bocquet, A. Noullez  and Z. R\'acz
for many useful discussions. This work was supported by the P\^ole Scientifique de Mod\'elisation
Num\'erique at the \'Ecole Normale Sup\'erieure de Lyon.}

\end{document}